%% file: IncagliLomonosov21.tex
\documentclass[%reprint,
 amsmath,amssymb,
 aps,
showkeys,
showpacs
]{revtex4-2}

\usepackage{xcolor}
\usepackage{xspace}
\usepackage{siunitx}
\usepackage{caption}
\usepackage{subcaption}
\usepackage{verbatim}
\usepackage{rotating}
\usepackage{bm}
\usepackage{ulem}
\usepackage{comment}
\usepackage{graphicx}% Include figure files
\usepackage{dcolumn}% Align table columns on decimal point
\begin{document}

\input{sections/commands}

\preprint{MUPB/Conference section: }

\title{Measurement of the muon magnetic anomaly \amu in the \\
Muon \gm  experiment at Fermilab}
%Insert your title here\\with Forced Linebreak}% Force line breaks with \\

\author{Marco Incagli}
 \email{marco.incagli@pi.infn.it}
\affiliation{%
Istituto Nazionale di Fisica Nucleare, INFN -- Pisa} 

\collaboration{Fermilab E989 Muon g-2 Collaboration}%\noaffiliation

\date{\today}% It is always \today, today,

\begin{abstract}
The Fermilab Muon g-2 experiment measures the muon anomalous magnetic moment with high precision. Together with recent improvements on the theory front, the first results of the experiment confirm the long-standing discrepancy between the experimental measurements and the Standard Model predictions. 
The observed value of $a_\mu({\rm FNAL}) = 116\,592\,040(54) \times  10^{-11}  ~ (\text{0.46\,ppm})$, combined with the previous experimental measurement, results in a 
discrepancy of $ (251 \pm 59)\times
10^{-11}$ with the theoretical prediction, corresponding to $4.2 ~\sigma$.
 This note presents the first results, the current status and the future prospects of the Muon g-2 experiment at Fermilab.
%An article usually includes an abstract, a concise summary of the work
%covered at length in the main body of the article. 
%\begin{description}
%\end{description}
\end{abstract}

%\pacs{Suggested PACS}% PACS, the Physics and Astronomy
\maketitle

%\tableofcontents

% 1. Introduction to g-2
%\input{sections/introduction}

% 2. Introduction to the g-factor
\input{sections/TheGFactor}

% 3. The muon g-2
\input{sections/TheExperiment}

% 4. wa term
\input{sections/Precession}

% 5. wp term
\input{sections/MagneticField}

% 6. result
\input{sections/result}

%7. acknowledgments
\input{sections/acknowledgement}

%bibliography
\bibliography{sections/Biblio}% Produces the bibliography via BibTeX.
\bibliographystyle{elsarticle-num}

\end{document}

%% file: sections/commands.tex
\newcommand*\diff{\mathop{}\!\mathrm{d}}
\newcommand*\Diff[1]{\mathop{}\!\mathrm{d^#1}}
\newcommand{\pb}{PbF$_2$}
\newcommand{\mtca}{$\mu$TCA}
\newcommand{\fix}[1]{{\color{red}{#1}}}
\newcommand{\needref}{[{\color{red}{ref}}]}
\newcommand{\gm}{\ensuremath{g-2}\xspace}
\newcommand{\wa}{\ensuremath{\omega_{a}}\xspace}
\newcommand{\wam}{\ensuremath{\omega_{a}^{m}}\xspace}

\newcommand{\opprime}{\ensuremath{\omega'^{}_p}\xspace}
\newcommand{\opprimetilde}{\ensuremath{\tilde{\omega}'^{}_p}\xspace}
\newcommand{\optilde}{\ensuremath{\tilde{\omega}^{}_p}\xspace}
\newcommand{\Rmu}{\ensuremath{{\mathcal R}_\mu}\xspace}
\newcommand{\Rmuprime}{\ensuremath{{\mathcal R}'^{}_\mu}\xspace}

% the \oa command is used much more heavily, commenting these out
%\newcommand{\wa}{\ensuremath{\omega_{a}}\xspace}
%\newcommand{\wam}{\ensuremath{\omega_{a}^{m}}\xspace}

\renewcommand{\wp}{\ensuremath{\omega_{p}}\xspace}
\newcommand{\amu}{\ensuremath{a_{\mu}}\xspace}
\newcommand{\oa}{\ensuremath{\omega_{a}}\xspace}

\newcommand{\op}{\ensuremath{\omega_{p}}\xspace}
\renewcommand{\ns}[1]{\SI{#1}{ns}\xspace}
\newcommand{\mus}[1]{\SI{#1}{\micro\second}\xspace}
\newcommand{\mum}[1]{\SI{#1}{\micro m}\xspace}
\newcommand{\runone}{Run-1\xspace}
\newcommand{\runtwo}{Run-2\xspace}
\newcommand{\runthree}{Run-3\xspace}
\newcommand{\runfour}{Run-4\xspace}
\newcommand{\runonea}{Run-1a\xspace}
\newcommand{\runoneb}{Run-1b\xspace}
\newcommand{\runonec}{Run-1c\xspace}
\newcommand{\runoned}{Run-1d\xspace}
\newcommand{\precession}{precession-run1}
\newcommand{\field}{field-run1}
\newcommand{\BD}{BD-run1}
\newcommand{\PRL}{PRL-run1}
\newcommand{\Tr}{\ensuremath{T^{}_{r}\xspace}}

\newcommand{\musec}{$\mu$sec\xspace}
\newcommand{\pe}{\ensuremath{\phi_{\rm ens}}\xspace}

%% file: sections/TheGFactor.tex
\section{The $g$-factor and the muon magnetic anomaly}
\label{sec:Gfactor}
The $g$ factor relates the magnetic moment of a particle to its
angular momentum and charge-to-mass ratio..  
For a charged lepton, $g$ relates its magnetic moment to its spin:
\begin{equation}
\vec{\mu} = - g \frac{e}{2m} \vec{S}.
\end{equation}
Experimentally, it was found that $g=2$, but only in 1928 this value was derived by Dirac starting from his famous formula. A spectacular success of the Quantum Theory.

The {\it magnetic anomaly} is the fractional difference of $g$ from
the value 2:  $a = \frac{g-2}{2}$. 
Experimental evidence that $g\ne 2$ began mounting by 1947 through measurements such
as the Lamb shift~\cite{PhysRev.72.241} and  preliminary
measurements of $g$ factors in gallium by Kusch and
Foley~\cite{PhysRev.72.1256.2} indicating an incomplete understanding
of electrodynamics at atomic scales.  These and other results drove  
%Schwinger~\cite{PhysRev.73.416}, 
Schwinger, Feynman, Tomonaga
and others to combine electromagnetism with the quantum theory and thereby provide the foundation of Quantum Electrodynamics (QED).
QED predicted the possibility for charged particles to emit and
reabsorb particles from the quantum vacuum, thus modifying the
effective coupling constants.  This manifestly quantum effect enhances
the $g$ factor to a value larger than 2, resulting in a non-zero
anomaly.  The famous Schwinger term, published in 1948, 
\begin{equation}
a = \frac{\alpha}{2\pi} \sim 0.00116,
\end{equation}
provides the leading contribution to the muon and electron magnetic
anomaly\footnote{Higher order contributions depend on the mass and
  thus are different for the two leptons}.
Earlier that same same year, Kusch and Foley~\cite{PhysRev.74.250}, studying the Zeeman effect in Gallium atoms, published their definitive measurement of a non-null value of the magnetic anomaly for the elctron, finding
\begin{equation}
a^{exp} = 0.00119 \pm 0.00005.
\end{equation}
Schwinger's prediction aligned in perfect agreement with the measurement and together they confirmed the existence of these {\it radiative corrections}.
Another important success of QED.

%Modern Feynman diagrams allow representation of the interaction of the
%magnetic moment of the muon with an external field as in
%Fig.~\ref{fig:feynman}. The diagram on the left represents the {\it
%  tree level} process, corresponding to $g=2$, while the one on the
%right provides the first correction due to the emission and
%re-absorption of a virtual photon. 
%Note that the first correction
%$a=\alpha/2\pi$ is the same for all leptons, while higher order
%corrections include mass terms and thus differ for the different
%particles. 

%\begin{figure}[htbp]
%\centering
%\includegraphics[width=0.5\linewidth]{figures/GM2Feynman.png}
%\caption{Feynman diagrams representing the muon interaction with an
%external magnetic field. Left: tree level. Right: first order
%correction due to vacuum polarization corresponding to the Schwinger
%term $\alpha/2\pi \sim 10^{-3}$.} 
%\label{fig:feynman}
%\end{figure}

Since then, many more diagrams contributing to \amu have been
evaluated. These include the theoretical {\it tour de force} of the QED
contributions to 5 loops (12,672 diagrams)
%~\cite{Aoyama:2012wj} 
and the important weak interaction contributions.
%~\cite{PhysRevD.67.073006,PhysRevD.73.119901,PhysRevD.88.053005}.   
Many efforts have contributed to the evaluation of the QCD
contributions
%~\cite{Davier:2017zfy,Keshavarzi:2018mgv,Colangelo:2018mtw,Hoferichter:2019gzf,Davier:2019can,Keshavarzi:2019abf,Kurz:2014wya,Melnikov:2003xd,Masjuan:2017tvw,Colangelo:2017fiz,Hoferichter:2018kwz,Gerardin:2019vio,Bijnens:2019ghy,Colangelo:2019uex,Blum:2019ugy,Colangelo:2014qya} 
in the report of the Muon g-2 Theory Initiative
(see~\cite{Aoyama:2020ynm} and references therein).  Their consensus
value of 
\begin{equation}
\amu^{theo} = 
(116,591,810 \pm 43) 
\times 10^{-11},
\end{equation}
corresponding to $370$ parts per billion (ppb), 
represents an impressive precision.

Similarly, the average of the published result \cite{PRL} by the E989 collaboration ({\it Fermilab \gm}) 
and the previous value published by the E821 collaboration~\cite{BNLFinal} at
Brookhaven National Laboratory (BNL) yields the experimental value, 
\begin{equation}
\amu^{exp} =  116, 592, 061 \pm 41) \times 10^{-11}
\end{equation}
corresponding to $350$ ppb.
Theory and experiment show a difference of $ (251 \pm 59)\times
10^{-11}$, which corresponds to $4.2$ standard deviations. This
difference can hide additional terms which are not accounted for by
the current Standard Model of Particle Physics.  
As discussed in section 6, a recent lattice calculation of the QCD
contribution to \amu ~\cite{Borsanyi:2020mff} reduces this discrepancy, but at the same time it
creates a tension with the value reported in~\cite{Aoyama:2020ynm}, therefore it is
going through a close scrutiny within the theoretical community. 

%\subsection{Why muons and not electrons?}
%The magnetic anomaly can be predicted theoretically and measured experimentally with a precision reaching almost two order of magnitudes better for electrons  than for muons. Why does the muon magnetic anomaly remain interesting?
%The reason lies in the fact that in a typical diagram due to Physics Beyond the Standard Model, like the one shown in Fig.~\ref{fig:BSM} for the exchange of a new hypothetical particle $Z'$, the intermediate vector boson line causes a flip of the chirality of the fermion. 
%This chirality flip suppresses the contribution by the second power of the mass, yielding a relative sensitivity of the magnetic anomaly to new physics for the muon versus the electron proportional to
%\begin{equation}
%    \frac{m_\mu^2}{m_e^2} \sim 40,000.
%\end{equation}
%This factor more than balances the loss in precision when measuring the anomaly for muons rather than for electrons.

%Of course specific theoretical models may have a different structure, thus this factor doesn't necessarily appear.

%\begin{figure}[htbp]
%\centering
%\includegraphics[width=0.3\linewidth]{figures/Gm2%-BSM.png}
%\caption{Typical Beyond the Standard Model %contribution to the magnetic anomaly in which a %hypothetical $Z'$ boson is exchanged between the %muon lines.}
%\label{fig:BSM}
%\end{figure}

%% file: sections/TheExperiment.tex
% !TEX root = ../main.tex
\section{The muon \gm  strategy}
\label{sec:MuonGm2}

\begin{figure}[tb]
\centering
\includegraphics[width=0.6\linewidth]{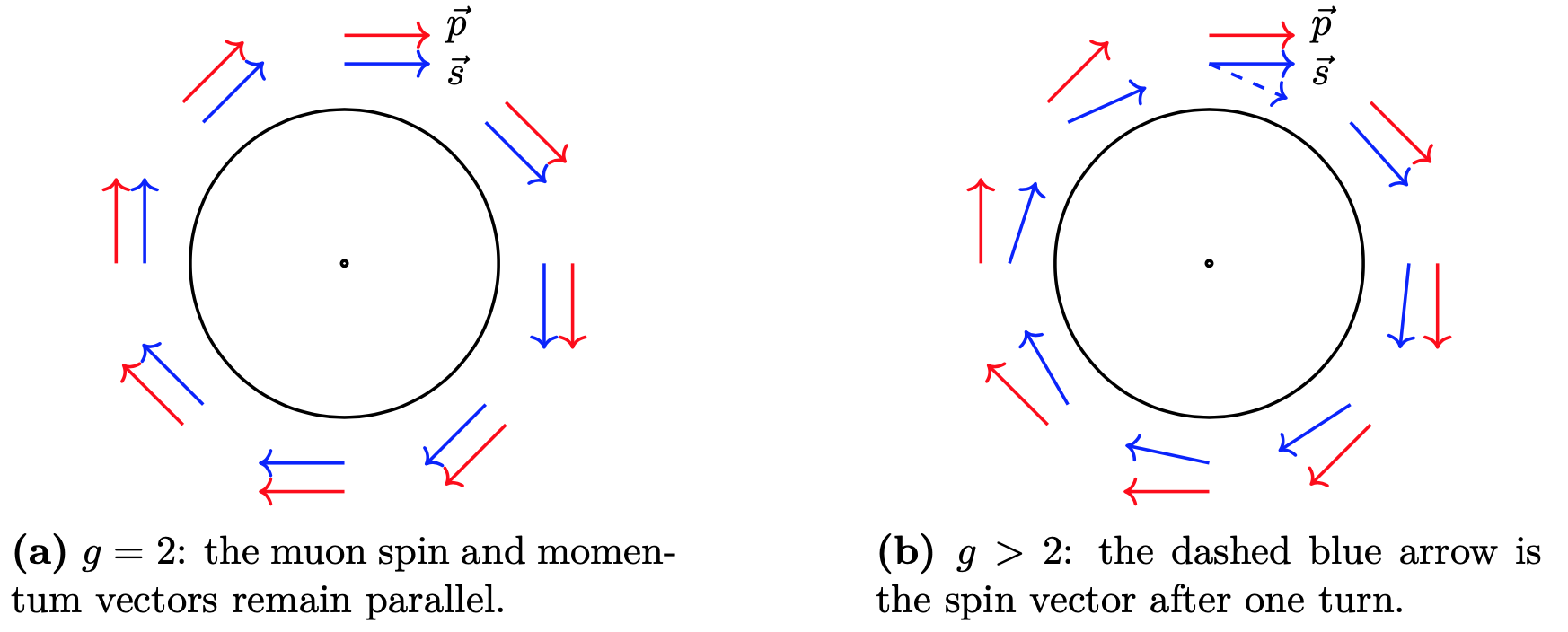}
\caption{Spin and momentum
  vectors for a muon orbiting in a magnetic field (a) when
  $a_{\mu}=0$, so the spin does not rotate relatively to the muon
  momentum, and (b) when $g>2$.} 
\label{fig:MomentumSpinMotion}
\end{figure}

\begin{figure}[tb]
\centering
\includegraphics[width=1.\linewidth]{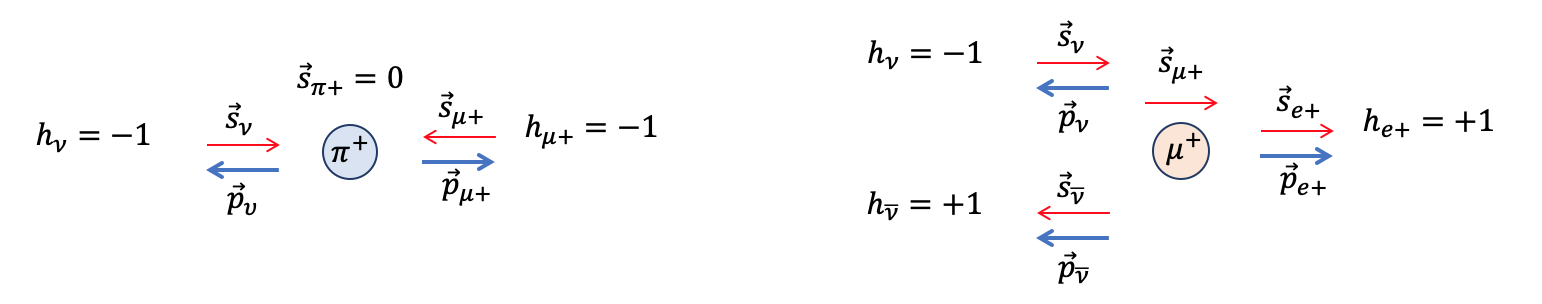}
\caption{Schematic representation of pion (left) and muon (right) decays. Blue arrows ($\vec{p}$) and red arrows ($\vec{s}$) represent the momentum and spin vector, respectively, while $h$ is the particle helicity.}
\label{fig:MuonPionDecays}
\end{figure}

The storage ring measurement of the muon anomaly relies on the spin precession and cyclotron motion of a charged particle orbiting in a uniform magnetic field.  
For a particle with momentum and spin vectors in a plane  perpendicular to $\vec{B}$, a classical calculation of the difference of these frequencies yields
%The technique used to measure exprimentally the muon anomaly relies on the behavior of a charged particle with spin immersed in a magnetic field. 
%Such a particle undergoes a cyclotron revolution while the spin precesses around the magnetic field. The difference between the angular velocities of these two motions can be simply evaluated in classical mechanics to be:
\begin{equation}
\oa = \omega_s - \omega_c = g\frac{e}{2m}B - \frac{e}{m}B =  \amu \frac{e}{m}B
\label{eq:oa}
\end{equation}
so that
\begin{equation}
\amu = \frac{\oa}{B} \frac{m}{e}
\label{eq:amu}
\end{equation}
A relativistic calculation modifies the expression for $\omega_s$ and $\omega_c$,  but the difference in Eq.~\ref{eq:oa} remains unaffected.
Thus, 
%assuming that at $t=0$ the momentum and spin vectors lie in a plane perpendicular to $\vec{B}$, if  
for $\amu=0$, that is $g=2$, the two vectors rotate with the same frequency, while for
%. On the other hand, if $g>2$, and thus 
$\amu > 0$, the spin vector rotates faster than the momentum vector (see
fig.~\ref{fig:MomentumSpinMotion}). 
In the Fermilab \gm setup, the spin advances by approximately
$12^o$ with respect to the momentum each orbit. 
%
%A measurement of \amu can be performed as follows:
%\begin{enumerate}
%\item produce a beam of polarised muons at $t=0$;
%\item have them evolve in a very stable and precisely measured magnetic field;
%\item measure the time and the spin direction at each muon's decay.
% \end{enumerate}
%
An observable sensitive to this relative precession rate would
therefore provide a direct measurement of $\amu$. This approach can be
realized using a beam of polarized muons that evolve in a very stable
and precisely measured magnetic field. 
Parity violation from the V-A structure of weak decays provides both a
source of polarized muons and a way to statistically identify the muon spin direction (see
fig.~\ref{fig:MuonPionDecays}). 
%Initial momentum selection of the
%$\pi^{+}$ just above the muon beam energy produces a beam with $96\%$
%polarization. 
%In muon decay, parity violation strongly correlates the highest energy
%positrons in the muon rest frame with the muon spin direction (see
%Fig.~\ref{fig:MuonPionDecays}). When coupled with the Lorentz boost,
%this spin-energy correlation results in a positron energy spectrum in
%the laboratory frame that depends on the muon spin  direction relative
%to its momentum. The spectrum varies from stiffest, when the
%directions align, to softest, when they anti-align. The spectrum thus
%modulates at the frequency \oa. 
 
%\subsection{Precession frequency}

Storage of the muon beam requires vertical focusing from a quadrupole
system, but the complicated spin precession in magnetic quadrupoles
would render precision measurement impossible. 
The experiment therefore employs electrostatic quadrupoles.
The electric field adds a $\vec{\beta}\times \vec{E}$ term,
corresponding to a $\vec{B}$ field in the muon rest frame, to the
expression in Eq.~\ref{eq:oa}.   
With an "out of plane" (vertical) momentum component also considered,
the spin evolves as~\cite{Jackson}: 
 \begin{equation}
 %\begin{aligned}
\frac{d(\hat{\beta}\cdot\vec{S})}{dt} 
= -\frac{q}{m}\vec{S}_{T}
\cdot
\left[a_\mu\hat{\beta}\times\vec{B} 
\right. 
\left.
+ \beta \left(a_\mu - \frac{1}{\gamma^{2}-1}\right)\frac{\vec{E}}{c} \right]
\label{eq:trueomega}
%\end{aligned}
\end{equation}
%\begin{eqnarray}
%    \vec{\omega}_a &=& -\frac{e}{mc}
%    \left [
%    \amu \vec{B} - \amu \left (
%    \frac{\gamma}{\gamma+1}
%    \right ) (\vec{\beta} \cdot \vec{B}) %\vec{\beta}
%    \right. \nonumber \\
%    & & \left. -
%    \left (
%    \amu - \frac{1}{\gamma^2-1}
%    \right )
%    \frac{\vec{\beta} \times \vec{E}}{c}
%    \right ]
%    \label{eq:oacomplete}
%\end{eqnarray}
where 
%\begin{equation}
$
    \vec{S}_T = \vec{S} - (%\left(
    \hat{\beta} \cdot \vec{S} )%\right)
    \hat{\beta}
%\end{equation}
$ 
is the spin component perpendicular to the momentum direction $\hat{\beta}$.
With $\vec{E}=0$ and the spin and momentum restricted to a plane perpendicular to $\vec{B}$, Eq.~\ref{eq:trueomega} reduces to the simple Eq.~\ref{eq:oa}.

%In the complete expression Eq.~\ref{eq:trueomega}, the second term represents the electric field contribution to the precession angular velocity. 

Farley, Picasso and collaborators \cite{CERN-Mainz-Daresbury:1978ccd} realized in the 70s that the strategic choice of
%\begin{equation}
%\gamma = \sqrt{\frac{\amu +1}{\amu}} \sim 29.3
%\end{equation}
$\gamma = \sqrt{(\amu +1)/\amu} \sim 29.3$
corresponding to a muon momentum $p_{0}=3.094$~GeV$/c$,
would minimize the electric field contribution to \oa.
At this {\it magic momentum}, the prefactor of the $\vec{E}$ term vanishes.  
Because of the 
finite Storage Ring momentum acceptance of
\begin{equation}
    \delta p/p = 0.15 \%,
\end{equation}
the cancellation occurs only at first order, but it allows treatment of the E-field contribution as a correction to the measured \oa. 

\paragraph{Utilizing comagnetometry}
Measurement of the magnitude of the field $|\vec{B}|$ by nuclear
magnetic resonance (NMR) probes, as detailed in the next section,
allows its expression in terms of the precession frequency of protons
shielded in water $\opprimetilde(T)$ as  
\begin{equation}
	\tilde{B}=\frac{\hbar \opprimetilde(T)}{2\mu'^{}_p(T)} =
        \frac{\hbar \tilde{\omega}'^{}_p(T)}{2}
        \frac{\mu^{}_e(H)}{\mu'^{}_p(T)}\frac{\mu^{}_e}{\mu^{}_e(H)}\frac{1}{\mu^{}_e}, 
	\label{eq:Btoomega}
\end{equation}
with the last three factors known precisely.  The tilde in $\tilde{B}$
and $\opprimetilde(T)$ indicates 
the average of the field over the muon distribution weighted by the
detected decays over time.  
%For a perfectly uniform dipole B-field, this average over the muon
%distribution would not be required, but at the ppm level additional
%multipoles are present, as detailed in the next section.
%The external factors  $\mu^{}_e(H)/\mu'^{}_p(T)$, $\mu_e/\mu_e(H)$ and $\mu_e$ are
% known within an uncertainty of 25 ppb (see details in \cite{PRL}.
% relates
%the magnetic moments
%of an electron bound in hydrogen to that of a proton shielded in a
%spherical water sample, and is measured to \SI{10.5}{ppb} at a water
%temperature $\Tr=\SI{34.7}{\celsius}$~\cite{Phillips:1975kx}. 
%The
%bound-state QED corrections that determine the magnetic moment ratio
%of the electron bound in hydrogen versus a free electron
%$\mu^{}_e(H)/\mu^{}_e$
%\cite{Mohr:2015ccw}, 
%and the electron magnetic moment $\mu^{}_e$ are known to a fraction
%of~\cite{Tiesinga:2021myr}. 
Combining 
Eqs.~\eqref{eq:oa}, \eqref{eq:Btoomega}, and $\mu^{}_e =
\frac{g^{}_e}{2} \frac{e}{m^{}_e} \frac{\hbar}{2}$ yields 
\begin{align}
	a^{}_{\mu} = \frac{\omega^{}_a}{\opprimetilde(T)}
        \frac{\mu'^{}_p(T)}{\mu^{}_e(H)}
        \frac{{\mu^{}_e(H)}}{\mu^{}_e} \frac{m^{}_{\mu}}{m^{}_e}
        \frac{g^{}_e}{2}.
\label{eq:amuCalc} 
\end{align}

% The ratio of the mass of the muon and the mass of the electron
 %$m^{}_\mu/m^{}_e$ is known to \SI{22}{ppb} from the measurement of
 %the hyperfine splitting of muonium~\cite{Liu:1999iz}.
 %Finally, the $g$ factor of the electron
 %$g^{}_e$ is known to \SI{0.28}{ppt}~\cite{Hanneke:2008tm}. 

The Muon g-2 experiment thus provides the ratio
\begin{equation}
    \Rmuprime = \frac{\oa \cdot (1+C)} {\opprimetilde \cdot (1+B)}
\end{equation}
as its primary experimental output, where $C$ and $B$ represent small
corrections to the measured frequencies,  
related to beam dynamics ($C$) and to the presence of transient fields
($B$) as discussed in the next two sections.

The external factors -- the ratio of the magnetic moment of a proton shielded in a spherical water sample at a reference temperature of $T=\SI{34.7}{\celsius}$ to the magnetic moment of an electron bound in hydrogen ($\mu'^{}_p(T)/\mu_e(H)$), the ratio $\mu_e(H)/\mu_e$, the ratio of the muon to the electron mass and the $g$ factor of the electron $g_e$ -- are known  with a combined uncertainty of 25 ppb  (see details in \cite{PRL}).

%The experimental output is thus the ratio between the two angular frequencies 
%\oa and \opprimetilde.
%
%Due to effects related to beam dynamics, this ratio can be written as:
%\begin{equation}
%    \Rmuprime = \frac{\oa \cot (1+C)} {\opprimetilde \cdot (1+B)}
%\end{equation}
%where the term $C$
% summarizes the corrections due to beam dynamics, like the ones discussed above related to the $\vec{E}$-field and to the {\it pitch} effect,
%while $B$ takes into account transient fields generated by the storage ring elements, kickers and quadrupoles in particular. 
%These corrections will be described in the next two Sections.

%% file: sections/Precession.tex
%!TEX root = ../main.tex
\section{Measuring the anomalous precession frequency}

\begin{figure}
\centering
\begin{minipage}{.5\textwidth}
  \centering
  \includegraphics[width=.7\linewidth]{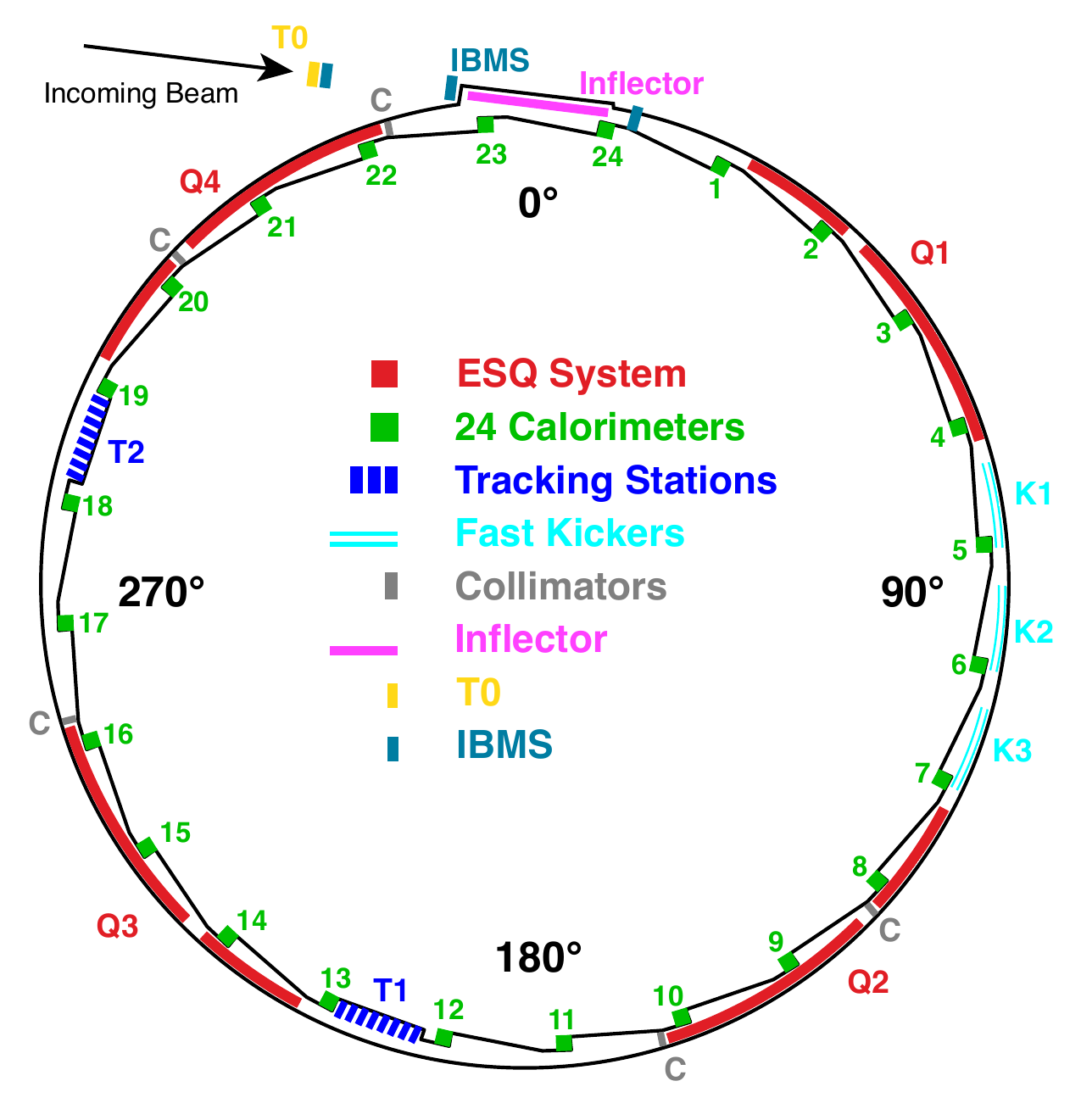}
 \captionof{figure}{Layout of the Muon g-2 experiment at Fermilab}
 \label{fig:layout}
\end{minipage}%
\begin{minipage}{.5\textwidth}
  \centering
  \includegraphics[width=1.1\linewidth]{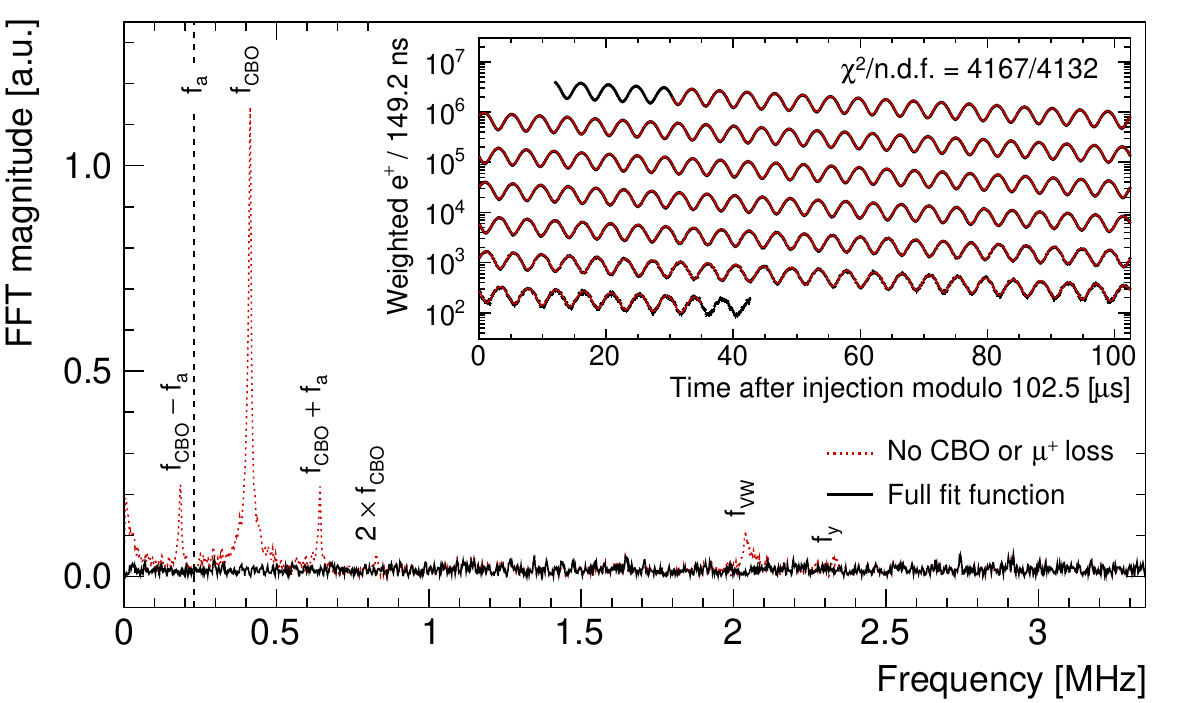}
  \captionof{figure}{Fourier transform of the residuals from a fit following Eq.~\ref{basicPrecession}.}
  \label{fig:fitPlots}
\end{minipage}
\end{figure}

The Fermilab complex delivers a sequence of 16 polarized muon bunches
every 1.4 seconds to the Muon g-2 storage ring, where each bunch
circulates for 700~\musec (a ``fill''), about 11 muon lifetimes. 
A suite of 24 PbF$_{2}$ crystal calorimeters~\cite{calo_refs} situated uniformly around
the interior of the storage ring (see Fig.~\ref{fig:layout}) detect
the positrons from beam muon decay.
Every
calorimeter consists of a $9\times6$ array of crystals, each with a
Silicon Photomultiplier (SiPM) photodetector.

% \begin{figure}[tb]
%\centering
%\includegraphics[width=.85\linewidth]{figures/RingDrawing_02272021_Full_degrees.pdf}
%\caption{Layout of the Muon g-2 experiment at Fermilab}
%\label{fig:layout}
%\end{figure}%

%To avoid bias by the data analyzers towards either the SM prediction
%or the previous measurements, the precise digitization rate remains
%unknown to the experiment until the the analysis has completed.  Two
%members of the Fermilab management team, not on the experiment, choose
%and monitor a frequency within a 50 part per million (ppm)
%``blinding'' range for each yearly run of the experiment. 

The variation of the positron energy spectrum as the spins in a
monochromatic polarized muon beam precess leads to a rate
time-dependence of the precession signal described by 
\begin{equation}
N(t) = N_{0}e^{-t/\gamma\tau_{\mu}} (1 + A(E_{thr}\cos{(\oa t + \phi_{\rm ens})} ),
\label{basicPrecession}
\end{equation}
where $\gamma$ is the standard boost factor (about 29.3 for muons at
the magic momentum), $\tau_{\mu}$ is the muon lifetime, \oa is the
anomalous precession frequency, and $A(E_{thr}$ is the asymmetry amplitude of
the sinusoidal variation, which depends on the energy threshold applied to the detected positrons.  The phase $\phi_{\rm ens}$ represents the
ensemble average precession phase for the muons with detected daughter
positrons.  That average phase receives several contributions: the
phase distribution within the injected beam, the longer drift distance
for higher energy positrons vs lower energy positrons because of their
different curvatures in the $\vec{B}$ field, and the detector
acceptance as a function of the transverse decay position of beam
muons. 
%The ensemble average phase \pe poses interesting challenges to the
%experiment.  
Any effect correlated with time after beam injection that
changes the makeup of muons with detected daughter positrons can lead
to a time dependent drift $\pe\rightarrow \pe(t) \sim
\phi_{0}+\phi_{1}t$.  The latter term would directly bias the value of
\oa extracted from the data.  
A rate-dependent drift of the gains, for
example, would change the effective energy threshold for detected
positrons and lead to such a drift.   
A laser system~\cite{laser_refs}
overlays well-characterized pulses on top of 10\% of our muon fills
that allow monitoring of and correction for such gain drifts. 
The
pileup of positrons close in time and space in a calorimeter, whose probability
varies as muons decay,
 can also
lead to such a drift.  
%In fact the detected low energy and high energy
%positrons correspond to systematically different muon phases 
%which,
%combined with the fact that
%the pileup rate varies as the beam intensity squared, and thus with a
%decay constant which is half the muon lifetime, would
%lead to bias on \oa if not corrected. 

%Approaching the reconstruction and analysis of precession signal with
%a variety of techniques provides valuable cross-checks of the the
%entire analysis process.  
The collaboration utilizes two complementary
techniques to reconstruct positron candidates from the waveforms,
which bring different optimizations for resolving pileup.  
%Five
%analysis groups employed three distinct data-driven approaches to
%correct their oscillation time series for the pileup.  
A third
technique reconstructs the total measured energy versus time, which
inherently eliminates bias from pileup.  All told six independent
analysis groups contributed 11 different measurements of \oa
(see~\cite{E989wapaper}).  
%A software package applied an additional
%blinding that differed for each group, shifting the results within an
%additional 50 ppm window by amounts unknown to any of the groups.
%Upon moving to a common blinding offset after all analysis groups had
%fully completed their analyses, including systematic uncertainty
%assessments, we found excellent agreement among the results. 

%Fits to the reconstructed time series determines \oa.  
Fitting with only the basic decay model of Eq.~\ref{basicPrecession}
results in set of residuals that show distinct frequencies in their
fast Fourier transform (FFT) shown in Fig.~\ref{fig:fitPlots}.  These
frequencies correspond to well-understood horizontal and vertical
oscillations of the stored beam particles about their nominal circular
orbits, which then couple to the acceptance of the detector system to
modulate the observed rates.  Appropriate modification of the basic
model to account for these effects results in excellent quality fits
that match the data well (see Fig.~\ref{fig:fitPlots}), have residuals
with a featureless FFT spectrum, and $\chi^{2}$ values consistent with
the number of degrees of freedom.  Combination of the four data
subsets in \runone, which correspond to different operating conditions,
provides an overall statistical precision of 434 parts per billion
(ppb). 

% \begin{figure}[tb]
%\centering
%\includegraphics[width=1.\linewidth]{figures/prl9dy.pdf}
%\caption{Fourier transform of the residuals from a time-series fit
%following Eq.~\ref{%basicPrecession} but neglecting betatron motion
%and muon loss (red dashed); and from t%he full fit (black). The peaks
%correspond to the neglected betatron frequencies and mu%n
%loss. Inset: Asymmetry-weighted $e^+$ time spectrum (black) from the
%Run-1c run grou%p fit with the full fit function (red) overlaid. 
%}
%\label{fig:fitPlots}
%end{figure}

\paragraph{Beam dynamics corrections}
The measured \oa value requires three significant corrections to allow
its interpretation as the frequency of Eqs.~\ref{eq:amu}
or~\ref{eq:amuCalc}.  The largest correction comes from the spread
of stored muon energies in the beam, which results in imperfect
suppression of the electric field term in Eq.~\ref{eq:trueomega}.  
%By
%analyzing the rate variation in the calorimeters as the momentum
%spread causes the beam to debunch, we can extract the momentum
%distribution and calculate the correction, which averages to 490 ppb
%over the four data subsets.  
A second correction results from vertical
momentum distribution of the beam muons, which alters the horizontal
precession rate.  A straw tracking system in the vacuum reconstructs
the beam motion by extrapolation of the decay positrons back to the
storage region.  
%The measured  vertical momentum distribution leads to
%corrections that average to 180 ppb.  
Finally, in \runone two faulty
high voltage resistors controlling the quadrupoles caused the beam to
change shape and to slowly drift downward during the time interval used to determine \oa. 
 When
coupled with acceptance effects, these changes resulted in a drift in
the ensemble average phase, thus biasing \oa.  
This effect has been
modelled and understood well.
%, resulting in corrections of -170 ppb. 

These corrections add up to a total shift $C \simeq 500$ ppb,
with an uncertainty of $93$ ppb, on the measured \oa value as reported
in the summary table~\ref{tab:syst}.  

%\fix{ Marco, not sure at this point how we want to summarize the results from the four data subsets}

%\fix{Results are in table 2. We will have to reduce quite substantially the text, and the figures, to match the limit. By looking at the last journal issues, the limit is not so stringent, but we cannot force too much. The trolley layout is probably superfluous. Maybe also 2.1 (muons vs electrons) can be reduced to a sentence.}

%% file: sections/MagneticField.tex
\section{The Magnetic Field \opprimetilde}

\label{sec:omega_p}
\begin{figure}
\centering
\begin{minipage}{.5\textwidth}
%\centering
\includegraphics[width=.7\textwidth]{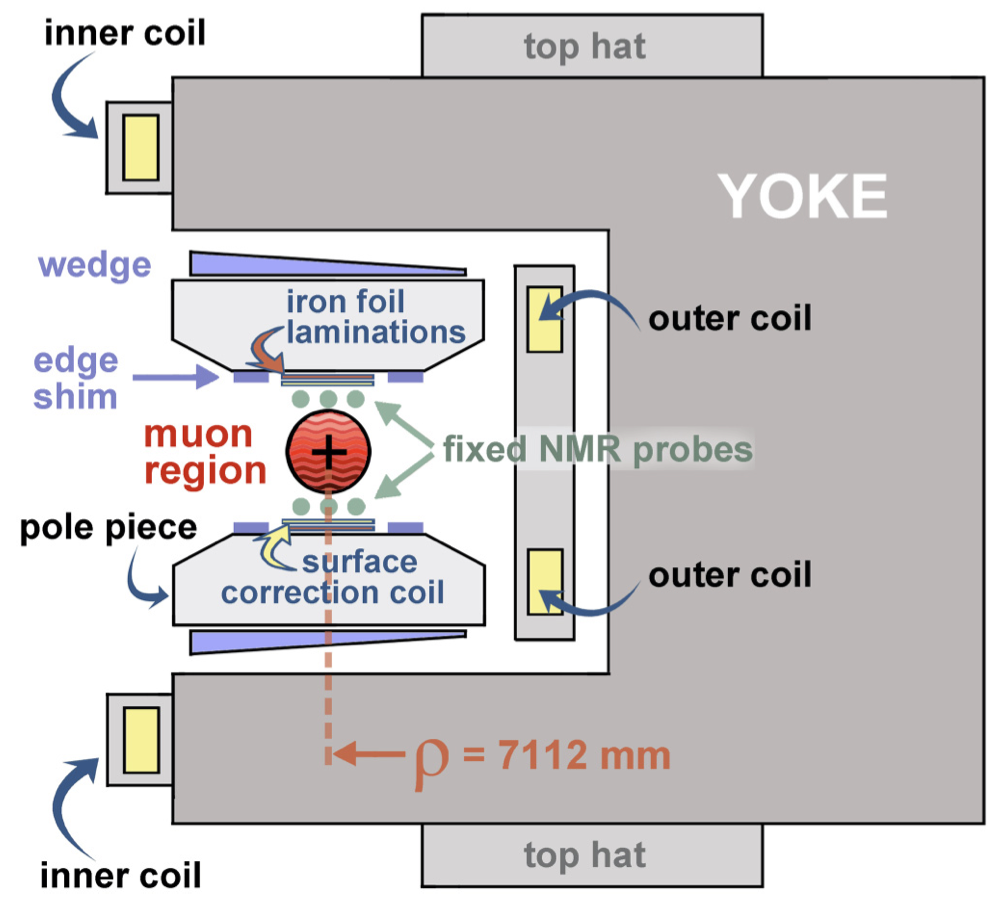}
\captionof{figure}{Cross section of the Muon \gm magnet. It's a
  $C$-shaped superconducting magnet that provides 
  a $1.45$~T field.} 
\label{fig:magnet}
\end{minipage}%
\begin{minipage}{.5\textwidth}
  %\centering
  \includegraphics[width=.8\linewidth]{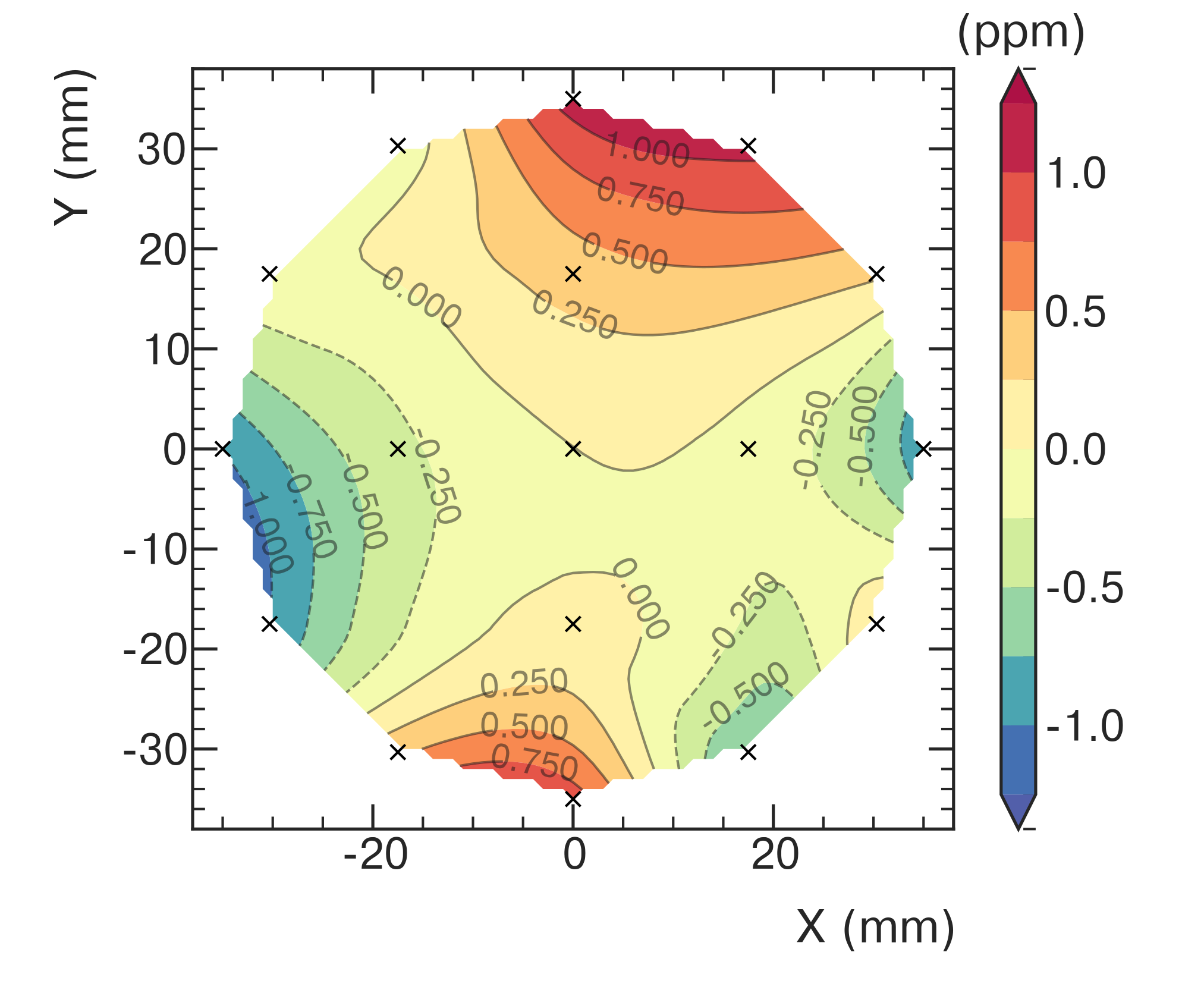}
  \captionof{figure}{Relative variation of the magnetic
      field.
      The locations of the 17 trolley
      probes are indicated by (x). }
  \label{fig:fieldMapTransverse}
\end{minipage}
\end{figure}

The $1.45$~T field is generated by a {\it C}-shaped superconducting
dipole magnet represented in figure~\ref{fig:magnet}. 
The magnetic
field in the $4.5$~cm radius storage region, described in detail
in~\cite{field-run1},  
is highly uniform in order
to reduce the uncertainty on the determination of the field
experienced by the muons. The uniformity is achieved by a long process
of shimming  that locally modifies the field direction.  
On top of this, an active feedback system modifies the coils
current in order to keep the magnetic field stable, for example for
hall temperature variations. 

\paragraph{Tracking the magnetic field}

The magnetic field is measured by using pulsed proton Nuclear Magnetic
Resonance (NMR) probes.
A cylindrically shaped {\it trolley}, which can run on rails inside
the storage region when muons are not present, 
hosts 17 NMR probes. 
%distributed as shown in Fig.~\ref{fig:TrolleySection}. 
%\begin{figure}
%    \centering
%\includegraphics[width=0.35\textwidth]{figures/TrolleySketch.png}
%\caption{The layout of the 17 probes in the trolley. Positive
%          $x$ is towards higher radius. The fixed probe locations on
%          the top (T) and bottom (B) of the storage region are shown
%          as well. For the fixed probes, the six-probe stations have
 %         probes in the inner (I), middle (M), and outer (O)
 %         positions. In the four-probe stations, only the middle and
 %         outer probes are present.} 
%	\label{fig:TrolleySection}
%\end{figure}
Each probe is filled with petroleum jelly and the
Larmor precession frequency of the protons within this jelly is
measured. Each probe is carefully calibrated in terms of a precision
calibration probe containing a pure water sample.
The in-vacuum trolley runs in the Storage Ring and measures the magnetic
field experienced by the muons in $\simeq 9000$ azimuthal locations. 
%The complete
%operation takes 3 hours, it requires to beam to be off and thus it is
%repeated every 3 days, in a compromise 
%between the precision of the measurement and the introduced dead time.

%Some stations
%host 6 and some 4 NMR probes located above and below
%the storage area as indicated in Fig.~\ref{fig:TrolleySection}.

%\begin{figure}[h]
%\centering
%\includegraphics[width=0.41\textwidth]{figures/magnet.png}
%\caption{Cross section of the Muon \gm magnet that provides the highly
%  uniform magnetic field required to achieve a such precise
% measurement. It's a $C$-shaped superconducting magnet that provides
%  a $1.45$~T field.} 
%\label{fig:magnet}
%\end{figure}
%\begin{figure}[!hbt]
%    \centering
%   \includegraphics[width=0.48\textwidth]{figures/fieldMapTransverse.png}
%    \caption{Relative variation of the azimuthally averaged magnetic
%      field with respect to the main dipole component.
%     The locations of the 17 trolley
%      probes are indicated by (x). } 
%   \label{fig:fieldMapTransverse}
%\end{figure}

The field's evolution between trolley runs is tracked by
 a  set of 378  probes which are mounted in 72
azimuthal stations regularly spaced around the ring. 
The measurement from the trolley probes
at a given azimuthal position $\theta$, is
determined by the solution of the source-free Laplace equation:
\begin{equation}
B = A_0 + \sum_{n=1} \left( \frac{r}{r_0} \right)^n 
\left[
A_n \cos{(n\theta)} + B_n \sin{(n\theta)}
\right]
\label{eq:moments}
\end{equation}
expressed in polar coordinates $(r,\theta)$ with respect to the center
of the muon ideal orbit, where $r_0=4.5$ cm is the radius of the
storage region.
The $A_n$ and $B_n$ parameters are the multipole strengths, also known
as the normal and the skew multipole, respectively.
The average over the azimuthal angle of the observed field, relative to
the dominant dipole component, is 
shown in Fig.~\ref{fig:fieldMapTransverse}, together with the location of
the measuring probes.
%Table~\ref{tab:multipoles} shows the %relative amplitude of the first
%multipoles with respect to the main %dipole.

%\begin{table}
%  \centering
%  \begin{tabular}{llrr} \hline \hline
%    \multicolumn{2}{c}{Moment %strength} & \multicolumn{1}{c}{normal} & \multicolumn{1}{c}{skew} \\
%    \multicolumn{2}{c}{[normalized to $A_0$]} & \multicolumn{1}{c}{[ppb]} & \multicolumn{1}{c}{[ppb]} \\
%    \hline
%    \rule{0pt}{1em}Dipole & $(A^{}_0,\ -)$ &  $10^9$ & - \\
%    Quadrupole & $(A^{}_1,\ B^{}_1)$&   300  &   399\\
%    Sextupole & $(A^{}_2,\ B^{}_2)$  & -1\,247 &   395\\
%    Octupole & $(A^{}_3,\ B^{}_3)$  &   14 &   273\\
%    Decupole & $(A^{}_4,\ B^{}_4)$  &   39 & -1\,319\\
%    Dodecupole & $(A^{}_5,\ B^{}_5)$ & -756 &  -187\\
%    Tetradecupole & $(A^{}_6,\ B^{}_6)$ & -1\,067 &    -0\\
%    \hline
%    \hline
%  \end{tabular}
%    \caption{Strength of each moment normalized to the strength
%      of the dipole $A^{}_0$ (see Eq.~\ref{eq:moments})
% averaged over azimuth.} 
%\label{tab:multipoles}
%\end{table}

The fixed probes are used to track the field  
in between trolley runs. 
%Since the number of probes per slice is limited, and the probes
%are outside the storage area, although still in vacuum, the tracking
%procedure 
%has an uncertainty which, in the first run of data taking, averaged to
%31 ppb among the different datasets. 

\paragraph{Calibration procedure}
The trolley probes are calibrated by means of an external probe 
hosting a
cylindrical water sample 
which is installed on a translation stage in the Storage Ring vacuum.
The translation stage allows the calibration probe to be moved to
each trolley probe position at a specific azimuthal location. 
%A set of local azimuthal coils is used to impose a known field
%gradient which allows a precise determination of the probe position
%with a precision of $\sim 0.5$ mm. 
The calibration and the trolley probes are then swapped several times
to obtain a calibration constant for each of the 17 probes.
%. The
%measurement time is $\sim 30$ seconds and the swapping time takes few
%minutes. 
%This procedure determines the calibration constants of individual
%trolley probes 
%with an average uncertainty of 30 ppb %which is
%dominated by the local magnetic perturbations from the magnetization
%of the materials used in the probes and trolley body.

\paragraph{Muon weighting}

The magnetic field map has to be averaged over the muon transverse
distribution at each azimuthal slice. The muon distribution is
measured at $\sim 180^o$ and $\sim 270^o$ with
respect to the injection point by two tracker stations.
%\cite{NIM-tracker}. 
The in-vacuum straw tracker stations measure the
trajectories of the decay positrons and trace them back to their
radial tangency point within the storage ring. 
These profiles are
propagated to other azimuthal locations using beam dynamics
simulation. 
%The uncertainty in the determination of the muon transverse
%distribution is dominated by the tracker 
%alignment and it is limited to 20 ppb.

\paragraph{Transient fields}
On top of the main static field, additional fields are induced by the
fast switching storage ring elements that define the muon trajectory:
the kicker and the electrostatic quadrupoles. 
 An eddy current induced locally %in the vacuum chamber structures 
 by the
 kicker system produces a transient magnetic field in the
storage volume. A magnetometer, installed between the kicker
plates, measures the Faraday rotation of a polarized laser
light 
in a terbium-gallium-garnet (TGG) crystal.
The second transient arises from charging the electrostatic quadrupoles,
where the Lorentz forces induce mechanical vibrations
in the plates that generate magnetic perturbations. 
Customized NMR probes
 measure these transient fields at several positions 
 %within
 %one ESQ and at the center of each of the other ESQs 
 to determine the average field throughout the quadrupole
 volumes.
%The two effects combine  to produce a total correction factor $B=-44$
%ppb  which is known with an estimated uncertainty of $99$ ppb (see
%table~\ref{tab:syst}). 

%% file: sections/result.tex
\section{Result and perspectives}
The recently published result~\cite{PRL} comprises four data subsets collected
between April and July 2018 with distinct beam storage conditions, and totals $10^{10}$ positrons in the analysis.
Table~\ref{tab:Rmu} lists the values of the muon and proton precession angular frequencies, \wa and \opprimetilde, for the four subsets along with 
the combined value for the ratio \Rmuprime. 
The systematic uncertainties correlate strongly among the four measurements, but the statitical term, which is uncorrelated among the subsets, dominates the total error.
Combining \Rmuprime with the external input in Eq.~\ref{eq:amu} yields a muon anomaly of
\begin{equation*}
a_\mu({\rm FNAL}) = 116\,592\,040(54) \times  10^{-11}  ~~~ (\text{0.46\,ppm}),
\end{equation*}

Table~\ref{tab:syst} summarizes the statistical and systematic contributions to the final result.
The observed \amu value is fully compatible with the previous BNL result, and combine to give an experimental average of
\begin{equation*}
\amu(\text{Exp}) = 116\,592\,061(41) \times 10^{-11}   ~~~(0.35\,\text{ppm}).
\end{equation*}

 \begin{table}
%\begin{ruledtabular}
\begin{tabular}{lccc}
Run & $\omega_a/2\pi$\,[Hz] & $\opprimetilde/2\pi$\,[Hz] & $\Rmuprime \times 1000$\\
\hline
1a & 229081.06(28) & 61791871.2(7.1) & 3.7073009(45)\\
1b & 229081.40(24) & 61791937.8(7.9) & 3.7073024(38)\\
1c & 229081.26(19) & 61791845.4(7.7) & 3.7073057(31)\\
1d & 229081.23(16) & 61792003.4(6.6) & 3.7072957(26)\\ \hline
&  &  & 3.7073003(17)\\
\end{tabular}
%\end{ruledtabular}
\caption{\runone group measurements of \wa, \opprimetilde, and their ratios \Rmuprime multiplied by 1000.}
\label{tab:Rmu}
\end{table}

\begin{table} [h]
%\begin{ruledtabular}
\begin{tabular}{lrr}
%0.8\textwidth}{lrr}
\hline\hline
Quantity & Correction (ppb) & Uncertainty (ppb)\\
% &(ppb) & (ppb)\\
\hline
\wam (statistical) & -- & 434\\
\wam (systematic) & -- & 56\\
\hline
$C$ & 500 & 93\\
\hline
%$f_{\text{calib}}
$\langle \omega_{p}^\prime(x,y,\phi)\times M(x,y,\phi)\rangle$ & -- & 56\\
$B$ & -44 & 99 \\ \hline
%$\mu'_p(34.7^\circ)/\mu_e$  & -- & 10\\
%$m_\mu/m_e$  & -- & 22\\
%$g_e/2$ & -- & 0\\
%\hline
%Total systematic & -- & 157 \\
Total external factors & -- & 25 \\
\hline
Totals &544 & 462\\
\hline\hline
\end{tabular}
%\end{ruledtabular}
\caption{Summary table of uncertainties and corrections.}
\label{tab:syst}
\end{table}

The E989 experiment has already collected over 10 times the statistics used for this first measurement, and continues to collect additional data with the goal of reducing the statistical error to $\sim 100$ ppb. 
The systematic uncertainty currently sits at $157$ ppb, a factor of 2 lower than in the previous BNL experiment. 
Work in progress should reduce this uncertainty down to the $\sim 100$ ppb level, which will allow E989 to reach its proposed total uncertainty goal of a $\sigma^{tot} \sim 140$ ppb, a factor of 4 more precise than the previous experimental result.

\paragraph{Discussion}

The new result confirms the value of \amu found previously by the BNL E821 experiment.
The new world average shows a discrepancy of $4.2$ standard deviations
with the theoretical prediction recommended by the Muon g-2 Theory
Initiative~\cite{Aoyama:2020ynm}.  Recent lattice QCD calculations,
and in particular a recent precise result from the BMW
collaboration~\cite{Borsanyi:2020mff}, hint at a smaller discrepancy
with the observed anomaly. This new prediction, however, is in tension
with the current one, which is based on a dispersion integral of
experimental $e^{+}e^{-}\rightarrow {\rm hadrons}$ cross section
measurements~\cite{Aoyama:2020ynm}.
As Ref.~\cite{Keshavarzi:2020bfy} notes, an increase
in the measured  hadronic cross section below $\sqrt{s} \sim 1$ GeV 
could reconcile the two predictions, although the required increase
would be an order of magnitude larger than the current experimental 
precision.
 Additional contributions above $\sim 1$ GeV are excluded
at the 95\% Confidence Level
as they result in tension with the prediction of
         fundamental parameters from the global electroweak fits,
         like the Higgs and W masses.
Because of this, the theory community continues to push  both
calculational approaches to test the compatibility of different
predictions in some detail.

%The possibility of observing a discrepancy from the Standard Model prediction offers many intriguing possibilities.
Should the current \amu prediction based on the dispersion integral
hold, and assuming the current experimental central value also holds,
the expected improvement in precision would ascertain the current
discrepancy of $251 \times 10^{-11}$ with an uncertainty in the $40-50
\times 10^{-11}$ range, which would provide strong evidence of physics
beyond the Standard Model (BSM physics).    
Such a discrepancy, of the same order of magnitude as the electroweak
contribution to \amu ($154 \times 10^{-11}$),  would indicate a TeV
scale for the BSM physics.   
Even if the prediction and experimental determination should agree in the
end, the improvement in \amu will provide a powerful constraint on any
model extending the Standard Model.   
The next few years will provide exciting opportunities as the Muon g-2
experiment and the theory community continue to push on this precision
frontier.

%The picture changes if some hidden systematics is found in the $e^+e^-$ prediction.
%As it has been pointed out (\cite{Keshavarzi:2020bfy}), an increase in the value of the hadronic cross section, in particular in the $2\pi$ contribution, below $\sqrt{s} = 1$ GeV of $\sim 4 \%$ could reconcile the prediction based on dispersion intergral with the BMW lattice result. 
%However, this would create a tension in the electroweak fit, in what the fine structure constant $\alpha(Q^2)$ can also be derived from a similar dispersion integral. 

%% file: sections/acknowledgement.tex
\section*{Acknowledgments}
The Muon \gm Experiment was performed at the Fermi National
Accelerator Laboratory, a U.S. Department of Energy, Office of
Science, HEP User Facility. Fermilab is managed by Fermi Research
Alliance, LLC (FRA), acting under Contract No. DE-AC02-07CH11359.
Additional support for the experiment was provided by the Department
of Energy offices of HEP and NP (USA), the National Science Foundation
(USA), the Istituto Nazionale di Fisica Nucleare (Italy), the Science
and Technology Facilities Council (UK), the Royal Society (UK), the
European Union's Horizon 2020 research and innovation programme under
the Marie Sk\l{}odowska-Curie grant agreements No. 101006726,
No. 734303, the National Natural Science Foundation of China
(Grant No. 11975153, 12075151), MSIP, NRF and IBS-R017-D1 (Republic of Korea),
the German Research Foundation (DFG) through the Cluster of
Excellence PRISMA+ (EXC 2118/1, Project ID 39083149).